\documentclass[twocolumn,9pt]{article} % here we use the article class, rather than elsarticle

\usepackage[square,numbers,sort&compress,comma]{natbib}

\usepackage{amsmath}
\usepackage{amssymb}
\usepackage{caption}
\usepackage{graphicx}
\usepackage{latexsym}
\usepackage{times}
\usepackage[pagewise]{lineno}
\usepackage{hyperref}
\usepackage{xcolor}

\topmargin - 12pt % might need to be set to 0pt for some installations
\renewenvironment{abstract}%
              {% - begin definition
               \small% - select font
               {\bfseries \abstractname}% - select font
               \par% - end a paragraph (skip \parsep)
               \vspace{10pt}% - add vertical space
              }% - complete definition

\renewcommand\abstractname{Abstract}

\newcommand{\nomenclature}% - name of command
              [1]% - number of arguments
              {% - begin definition
               \bgroup% - begin a local group
               \flushleft% - turn on flushleft option
               \small\bf% - select font
               #1% - insert title text
               \par% - end a paragraph (skip \parsep)
               \egroup% - terminate local group
              }% - complete definition

\renewcommand{\section}% - name of command
              [1]% - number of arguments
              {% - begin definition
               \bgroup% - begin a local group
               \flushleft% - turn on flushleft option
               \small\bf% - select font
               \refstepcounter{section}% - increment counter
               \arabic{section}. #1% - insert title text
               \par% - end a paragraph (skip \parsep)
               \egroup% - terminate local group
              }% - complete definition

\renewcommand{\subsection}% - name of command
              [1]% - number of arguments
              {% - begin definition
               \bgroup% - begin a local group
               \flushleft% - turn on flushleft option
               \small\em% - select font
               \refstepcounter{subsection}% - increment counter
               \arabic{section}.% - insert title text
               \arabic{subsection}. #1% - insert title text
               \par% - end a paragraph (skip \parsep)
               \egroup% - terminate local group
              }% - complete definition

\renewcommand{\subsubsection}% - name of command
              [1]% - number of arguments
              {% - begin definition
               \bgroup% - begin a local group
               \flushleft% - turn on flushleft option
               \small\em% - select font
               \refstepcounter{subsubsection}% - increment counter
               \arabic{section}.% - insert title text
               \arabic{subsection}.% - insert title text
               \arabic{subsubsection}. #1% - insert title text
               \par% - end a paragraph (skip \parsep)
               \egroup% - terminate local group
              }% - complete definition

  \newcommand{\acknowledgement}% - name of command
              [1]% - number of arguments
              {% - begin definition
               \bgroup% - begin a local group
               \flushleft% - turn on flushleft option
               \small\bf% - select font
               #1% - insert title text
               \par% - end a paragraph (skip \parsep)
               \egroup% - terminate local group
              }% - complete definition

  \newcommand{\sectionbib}% - name of command
              [1]% - number of arguments
              {% - begin definition
               \bgroup% - begin a local group
               \flushleft% - turn on flushleft option
               \small\bf% - select font
               #1% - insert title text
               \par% - end a paragraph (skip \parsep)
               \egroup% - terminate local group
              }% - complete definition

\begin{document}

% -------------------------------------------------------------------- %
% -------------------------------------------------------------------- %
% -------------------------------------------------------------------- %

% -------------------------------------------------------------------- %

\small
\baselineskip 10pt

% -------------------------------------------------------------------- %
% -------------------------------------------------------------------- %
% -------------------------------------------------------------------- %
\setcounter{page}{1}
% -------------------------------------------------------------------- %
\title{\LARGE \bf Cavity-Stabilized Rotating Flames in a Circular Hele-Shaw Burner}

\author{{\large Xiangyu Nie$^{a}$, Shengkai Wang$^{a,*}$}\\[10pt]
        {\footnotesize \em $^a$ SKLTCS, CAPT, School of Mechanics and Engineering Science, Peking University, 5 Yiheyuan Road, Beijing, 100871, China}}

\date{}  %%% Leave as is, do not add date;

% -------------------------------------------------------------------- %
% -------------------------------------------------------------------- %
% -------------------------------------------------------------------- %
\twocolumn[\begin{@twocolumnfalse}
\maketitle
\rule{\textwidth}{0.5pt}
\vspace{-5pt}

\begin{abstract} % 100 to 300 words.
We report direct experimental observations of self-organized rotating flames of premixed methane (CH$_4$) and air in an open circular Hele-Shaw burner equipped with an annulus cavity flame holder. Unlike flames propagating in closed micro-channels, these flames exhibited stable traveling-wave patterns along the azimuthal direction of the cavity, where rapid expansion created a low-speed zone that facilitated flame stabilization. At low flow rates, the rotating flames were single-headed, with their rotation frequencies roughly proportional to the laminar flame speeds, suggesting that the flame fronts traveled in a nearly constant-shape fashion. As the flow rate increased, the rotating flames further split into multiple heads at approximately equal spacing, and the number of heads and rotation frequency generally increased with the flow rate, until these rotating flames transitioned into steady ring-shaped flames anchored at the cavity leading edge. Blow-off or extinction occurred at sufficiently high flow rates, where the flame front was pushed out of the rear side of the cavity. A series of parametric measurements were conducted over a wide range of equivalence ratios and flow rates, from which a regime diagram of different flame modes and their transition boundaries was obtained. Additional experiments were conducted on propane (C$_3$H$_8$) and dimethyl ether (DME) as well. It was found that the critical total mass flow rate at the rotating-steady flame transition boundary (approximately 10 SLPM) is insensitive to equivalence ratio, gap distance, and fuel type. A possible explanation for the invariance of the rotation-to-steady flame transition boundary was postulated based on a non-dimensional analysis involving a critical Peclet number. These results should be useful not only for the fundamental understanding of flame dynamics in micro-channels but also for the practical design of micro-combustors and the application of micro-combustion technologies.
\end{abstract}

\vspace{10pt}

{\bf Novelty and significance statement}

\vspace{10pt}
This work, to the best of the authors' knowledge, reports the first direct experimental observation of self-organized rotating flames inside a circular Hele-Shaw burner under both lean and rich conditions and without external heating. The rotating flame pattern, stabilized by an annulus cavity flame holder, accommodates a wide range of flow velocities and equivalence ratios through self-adjustment of the rotation frequency and wave number. The results are useful to the development of micro-combustors and for safety protection against unwanted fires and explosions in confined spaces. While recent studies of self-sustained rotating flames in Hele-Shaw cells show promise for modifying the effective burning rate beyond the limits of steady flames, the existence conditions of such rotating flames were largely uncertain. This study fills an important void by identifying a robust pattern of rotating flames, explaining its underlying mechanism, and presenting a regime diagram to facilitate combustion control.

\vspace{5pt}
\parbox{1.0\textwidth}{\footnotesize {\em Keywords:} Laminar flame propagation; Hele-Shaw burner; Traveling wave; Mode transition; Cavity flame holder}
\rule{\textwidth}{0.5pt}
*Corresponding author.
\vspace{5pt}
\end{@twocolumnfalse}]

% \linenumbers
\section{Introduction\label{sec:introduction}}\addvspace{10pt}

Understanding the propagation dynamics of flames in micro-channels is particularly important for the development of micro-combustion technologies \cite{maruta2011micro, ju2011microscale, FERNANDEZPELLO2002883} and for safety protection against unwanted fires and explosions in confined spaces \cite{martinez2019role}. A key feature of these flames is wall-induced thermal and chemical quenching, which significantly modifies their dynamics and instability characteristics compared with freely propagating flames. These effects can be further amplified by geometric confinement of the flow in narrow channels.

Historically, the propagation dynamics of flames in static mixtures within closed or semi-closed channels have been frequently studied. These flames have served as a platform to investigate various forms of combustion instabilities (for example, hydrodynamic \cite{al2019darrieus, shen2019flame, sarraf2018quantitative}, thermal-diffusive \cite{daou2021effect}, and thermal-acoustic \cite{veiga2019experimental} instabilities) under two-dimensional, simplified geometries. By contrast, studies of flames in flowing micro-channels have been comparatively scarce, partly because of the difficulty of designing devices that simultaneously allow well-controlled reactant supply, strong confinement, and optical access.

This issue can be resolved by using open Hele-Shaw burners with a central inlet. An exemplary series of work was conducted by researchers at Tohoku University \cite{kumar2007formation, kumar2007pattern, fan2009experimental, fan2013flame}, who used a burner-heated circular Hele-Shaw cell with well-controlled temperature profiles to investigate various flame patterns and their transition dynamics. Under external heating, the wall quenching effects were inhibited, enabling sustained propagation of the flame within the interior of the cell. In particular, an interesting pattern of rotating flames (spiral-like \cite{kumar2007pattern} or Pelton-like \cite{fan2009experimental}) was observed at average wall temperatures around 800 K. These self-sustained traveling flames are intriguing as they can modify the effective burning rate and total thermal power beyond the limits of their stationary counterparts, in a manner similar to rotating detonation waves in pressure-gain combustion systems \cite{zhou2016progress}. 

However, the exact range of conditions under which such rotating flames can exist remains largely unknown. In particular, it would be interesting to determine whether flame rotation can be sustained in the absence of external heating, i.e., in cold/unheated Hele-Shaw cells. From an engineering standpoint, the absence of continuous external heating is important because it makes the operating regime more relevant to practical compact combustors subject to strong heat losses.

In a very recent study, we reported a stable pattern of rotating flames formed along the side edge of an unheated Hele-Shaw burner \cite{Nie2026Hele}. Those flames were observed only under fuel-rich conditions, where the excess fuel created a traveling doublet structure, with a partially premixed branch gliding along the burner edge and a fully premixed branch extending into the interior. Reaction with the ambient air outside the burner played a critical role in the formation of such rotating flames. However, it remains uncertain whether a rotating flame could be sustained completely inside an unheated Hele-Shaw burner, especially under fuel-lean conditions. This is particularly important because lean-premixed combustion in confined space is widely pursued to improve efficiency and reduce emission in practical applications, whereas the lower burning velocity and stronger wall heat loss make flame stabilization substantially more difficult \cite{wan2017dynamics, veeraragavan2015flame, huang2009dynamics}.

The present study revisits the rotating-flame phenomenon with a modified burner configuration that confines the flame to the interior of the Hele-Shaw burner. Inspired by cavity-based flame holders, which have long been used to enhance flame stability by generating recirculation and low-velocity regions in combustors \cite{ben2001cavity, WANG20132073, wan2015experimental, YANG2026114807}, we introduce a miniature annular cavity located midway along the burner radius. This cavity creates a local low-speed zone through rapid flow expansion and effectively traps the flame inside the burner, thereby establishing an isolated environment for directly studying rotating-flame formation and transitions between different flame modes. Using this platform, we determine the conditions for the occurrence of rotation flames and explore the transitions mechanisms between rotating, steady, and locally extinguished flame states.

\section{Methods \label{sec:methods}} \addvspace{10pt}
\subsection{Experimental Method} \addvspace{10pt}

\begin{figure*}[h!]
\centering
\includegraphics[width=0.8\linewidth]{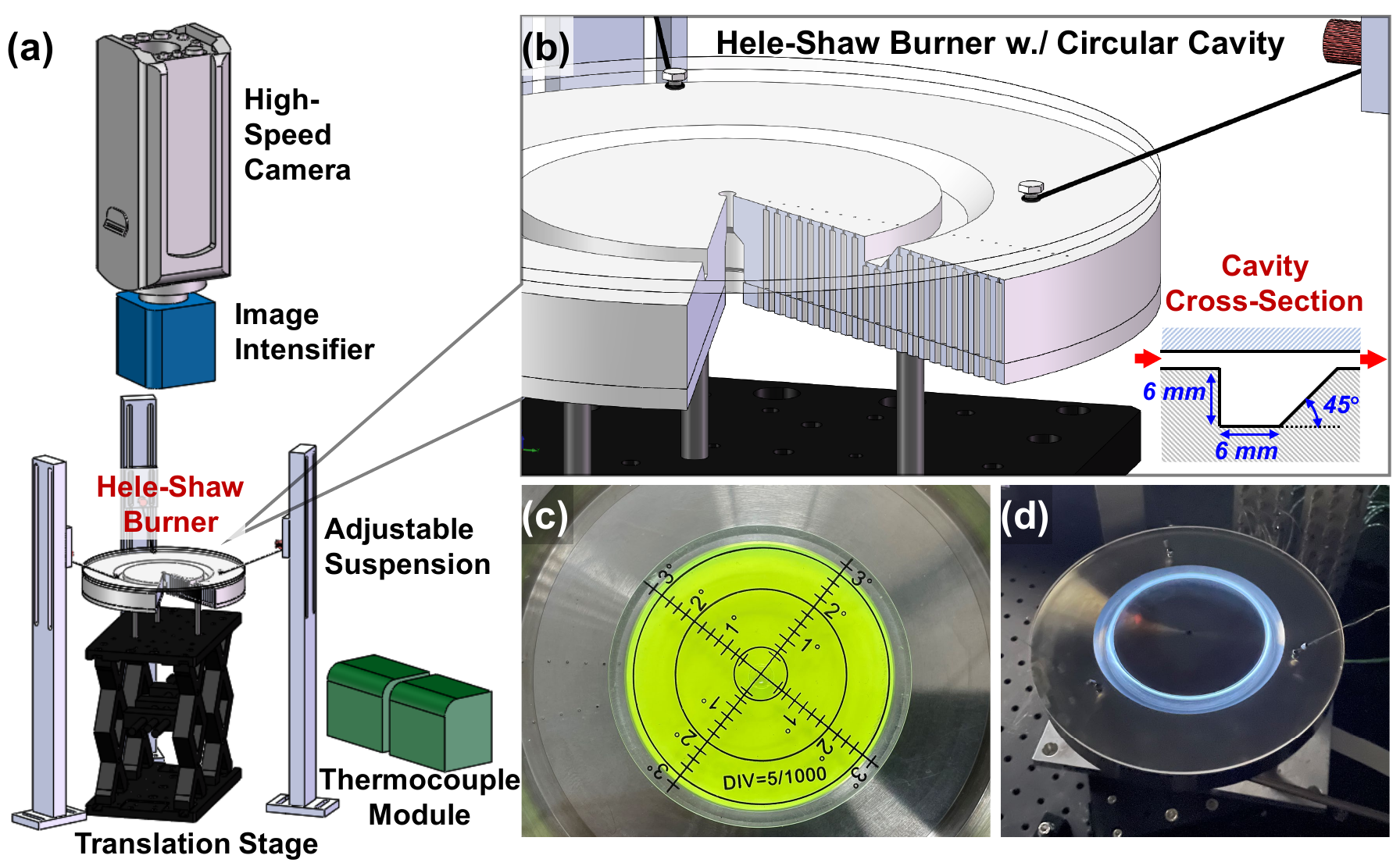}
\caption{\footnotesize Schematic of the current experimental setup. (a) Overall configuration of the burner and the optical diagnostic system. (b) Detailed view of the Hele-Shaw burner. (c) Horizontal angle of the top plate ($<0.1^{\circ}$) measured by a bubble level. (d) A long-exposure photo of a representative rotating flame in the burner cavity.}
\label{fig_01}
\end{figure*}

The flame experiments were conducted in a 200-mm-diameter circular Hele-Shaw cell formed by two parallel plates, as shown in Fig. \ref{fig_01}. The top plate was 2.5 mm thick and made of JGS1 fused quartz, enabling optical access from 185 to 2500 nm. The bottom plate was made of stainless steel with a thickness of 25 mm. The upper quartz plate was suspended by three thin wires and aligned with the bottom plate to an accuracy of 0.1$^{\circ}$ (as measured by a bubble level), ensuring a uniform gap distance across the burner. A cross-sectional view of the annulus cavity is shown in Fig. 1(b). The cavity was 6 mm deep and 6 mm wide at the bottom, with a 45-degree rear slope to suppress reflection of the impinging flow and acoustic waves. The leading edge of the cavity was located at a radial position of 49 mm, approximately halfway across the radius of the Hele-Shaw burner. The area of the cavity cross-section is 54 mm$^2$ (excluding the gap between the two plates).

The test gas mixture of high-purity CH$_4$ (99.99\%-grade) and synthetic air (prepared from 99.999\%-grade O$_2$ and 99.999\%-grade N$_2$) was supplied to the Hele-Shaw cell through a 4-mm-diameter entrance port at the center of the bottom plate. The fuel and air were thoroughly premixed in an in-line static mixer upstream of the entrance port, and their flow rates were accurately controlled by two Alicat MC-series mass flow controllers, with typical uncertainties of $\pm$ 0.1\% and $\pm$ 0.2\%, respectively.

To accurately measure the wall temperatures, two arrays of shielded K-type thermocouples with 0.5-mm tip diameter were embedded on the bottom plate of the Hele-Shaw burner. The thermocouples were flush-mounted on the upper surface of the plate to minimize disturbance to the flames. A total of 44 thermocouples were arranged along two radial lines separated by 15 degrees, providing spatially resolved temperature measurements from 12 to 98 mm along the radial direction at 2-mm intervals. The thermocouple signals were continuously acquired, digitally recorded at a sampling rate of 1 Hz, and streamed to a computer in real time via the RS-485 communication protocol using three 16-channel data acquisition modules (Rise PK9019). The uncertainty in surface temperature measurements was estimated to be $\pm$ 1\%, based on the thermocouple ratings.

The dynamic evolution of flame structure was continuously monitored using OH* chemiluminescence diagnostics. The OH* signal was spectrally filtered over 300 - 320 nm and recorded at a rate of 5 kHz using an image-intensified high-speed CMOS camera (Phantom v611 with EyeiTS intensifier). The recorded images had a pixel resolution of 512 $\times$ 512, with each pixel corresponding to a physical size of 280 $\mu$m $\times$ 280 $\mu$m. The effective spatial resolution, typically limited by the image intensifier, was determined to be approximately 0.3 mm (about 3 pixels) using a geometry calibration target. Further details on the calibration of the current imaging system can be found in the authors' previous work \cite{wang2025self}. 

For each experiment, the reactive mixture was first ignited at the burner edge at low flow rates, establishing a cavity-stabilized flame. The target operating condition was reached by gradually adjusting the flow rates. Note that the direction of flame rotation exhibited a degree of hysteresis. When a rotating flame first formed, it had a 50/50 probability of adopting either rotation direction. Once a stable pattern of flame rotation was established, it proved resistant to small perturbations in the flow rate or equivalence ratio. Measurements were taken after the flame pattern became stable.

\subsection{Numerical Method}
\addvspace{10pt}
 
Numerical simulations of cavity-stabilized flames were conducted using the EBIdnsFOAM solver \cite{bockhorn2012implementation, zirwes2023assessment} within the OpenFOAM framework \cite{weller1998tensorial}. Methane-air combustion was modeled using finite-rate chemistry with the DRM-19 kinetic mechanism, comprising 21 species and 84 elementary reactions  \cite{kazakov1994reduced}.

\begin{figure}[h!]
\centering
\includegraphics[width=\linewidth]{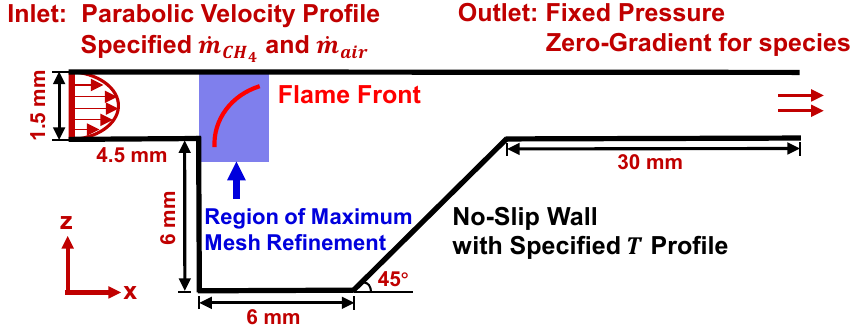}
\caption{\footnotesize \color{red}The cross-section \color{black}of the current computational setup for the two-dimensional axisymmetric simulations, with the region of maximum mesh refinement highlighted in blue.}
\label{fig_02}
\end{figure}

Two-dimensional axisymmetric wedge simulations were used to examine the radial flame stabilization mechanism of the steady ring-shaped flame. As shown in Fig.~\ref{fig_02}(a), the Hele-Shaw burner is represented numerically by an axisymmetric wedge sector of 1-degree angle. The computation domain spans $46.5~\mathrm{mm}$ in the axial flow direction ($x$) and $7.5~\mathrm{mm}$ in the vertical direction ($z$), comprising a main channel of $1.5~\mathrm{mm}$ height and a recessed cavity of 6 mm depth. The numerical geometry of the cavity is consistent with the experimental setup, featuring a vertical upstream wall, a 6-$\mathrm{mm}$ horizontal floor, and a 45-degree downstream ramp. The computation domain is discretized with a structured multi-block mesh, with local refinement ($\Delta x = \Delta z \approx 0.05~\mathrm{mm}$) near the cavity entrance where the flame is stabilized. Away from this region, the mesh is gradually coarsened toward the far-field boundaries, reaching a maximum cell size of approximately $1.0~\mathrm{mm}$. To ensure numerical stability and accuracy, the expansion ratio between adjacent cells is strictly limited to within $5\%$. The entire mesh consists of approximately $2 \times 10^4$ cells. Grid independence is validated by varying the grid size by factors of 0.5, 0.75, 2 and 4, with no observable change in the results.

A three-dimensional simulation was also conducted for a single-headed rotating flame in the cavity. The computational domain was constructed by extending the cavity cross-section over a 360-degree annular geometry. Adaptive mesh refinement (AMR) \cite{RETTENMAIER2019100317} was adopted to reduce the computational cost while maintaining sufficient spatial resolution near the propagating flame front. The mesh was dynamically refined around the reaction zone and coarsened in regions away from the flame. The AMR operation was performed every ten time-marching steps, using the OH mass fraction $Y_{\mathrm{OH}}$ as the refinement indicator. In particular, cells satisfying $Y_{\mathrm{OH}}>0.002$ were refined. The initial background mesh had a characteristic cell size of $0.5~\mathrm{mm}$, and three levels of refinement were used, resulting in a finest grid spacing of approximately $0.05~\mathrm{mm}$ near the flame front.

For both types of simulations, the inlet boundary condition is defined by a parabolic velocity profile based on the experimental values of total mass flow rates and mass fractions. The wall boundaries are defined as no-slip surfaces with experimentally determined temperature profiles. The outlet is assigned a boundary condition of constant pressure and zero concentration gradient.

\section{Results and Discussions \label{sec:results}}
\addvspace{10pt}

\subsection{Observation of Different Flame Modes}
\addvspace{10pt}

\begin{figure}[h!]
\centering
\includegraphics[width=\linewidth]{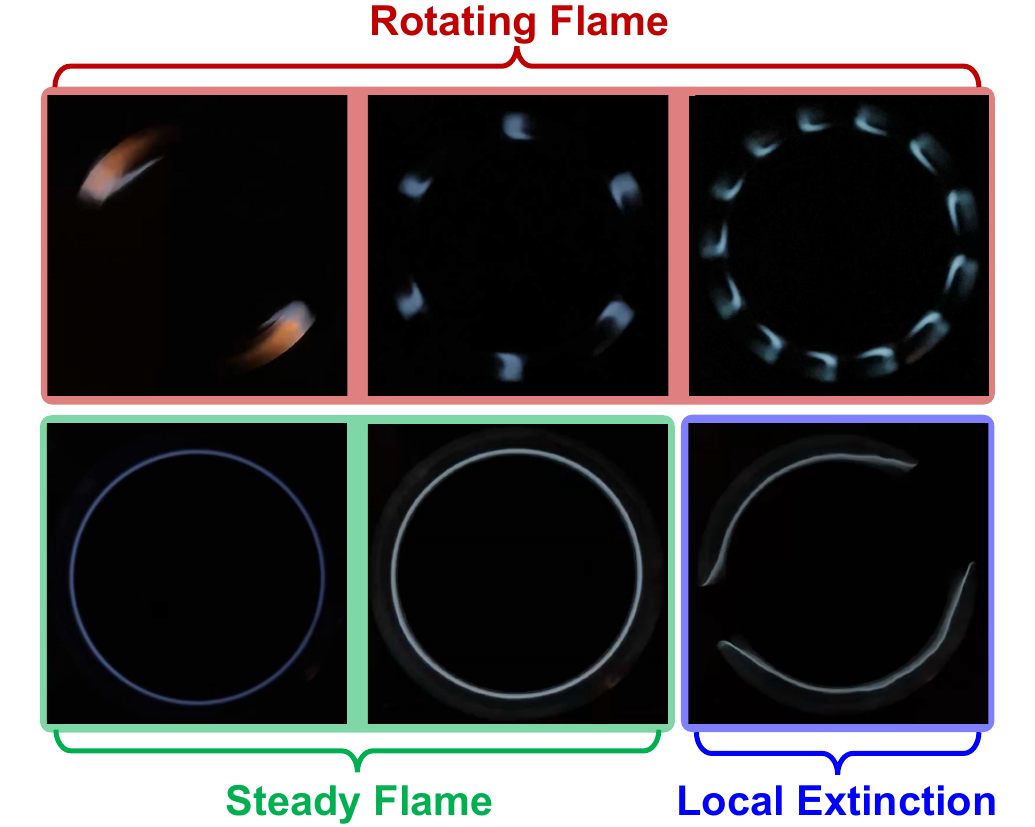}
\caption{\footnotesize Representative examples of three distinct flame modes. Top panels: rotating flames observed at relatively low flow rates; bottom left: steady ring-shaped flames at intermediate flow rates; bottom right: flames with local extinction at elevated flow rates.}
\label{fig_03}
\end{figure}

\begin{figure}[h!]
\centering
\includegraphics[width=0.9\linewidth]{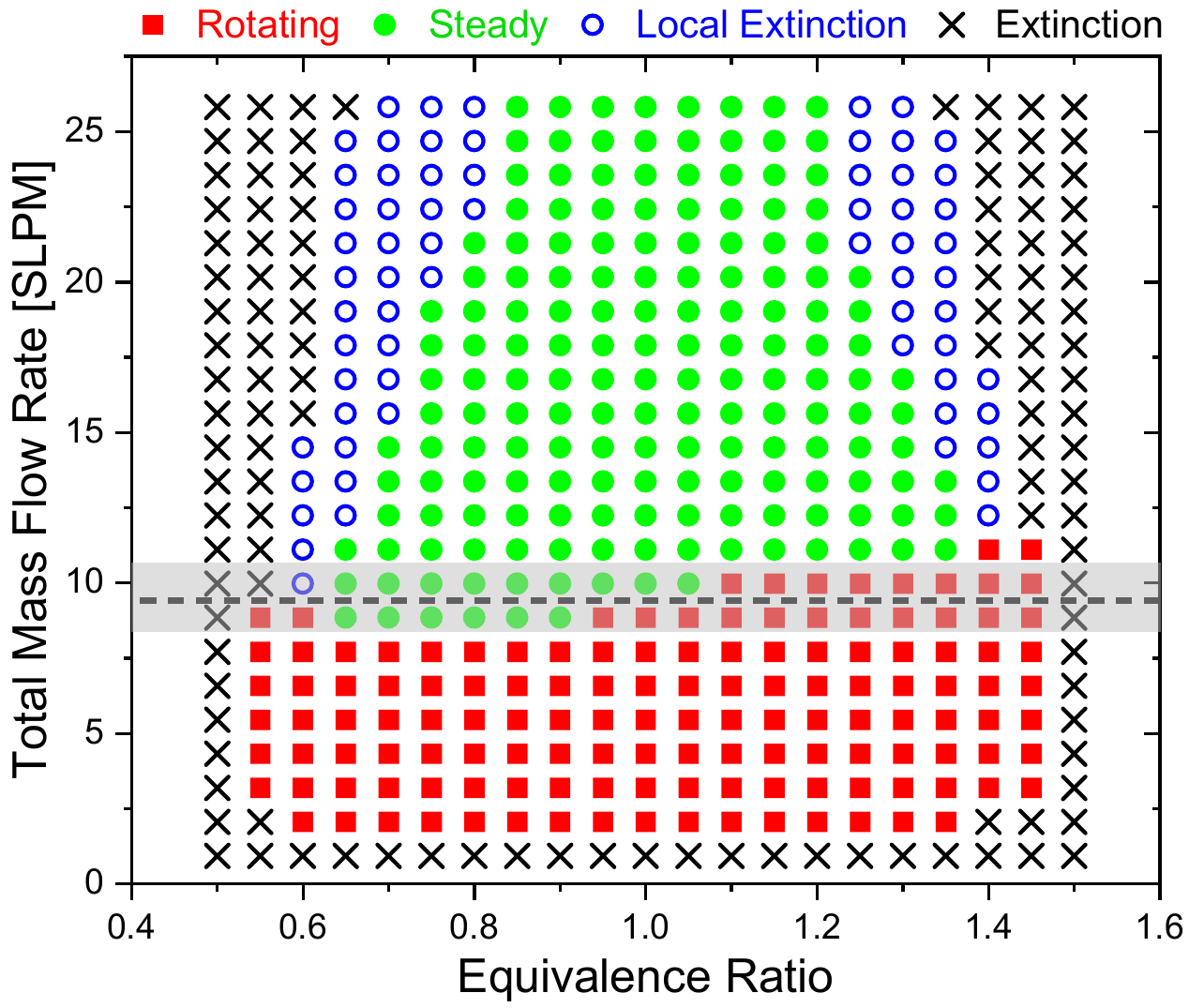}
\caption{\footnotesize Regime diagram of the experimentally observed flame modes at a gap distance of 1.5 mm. The dashed line indicates the transition boundary between rotating and steady flames, and the shaded band represents the associated uncertainty.}
\label{fig_04}
\end{figure}

Three distinct flame modes were observed over a range of equivalence ratios ($\phi$) and total mass flow rates ($\dot{V}$), as illustrated in Fig. \ref{fig_03}. At low flow rates, self-organized rotating flames formed spontaneously inside the cavity under both lean and rich conditions. These flames exhibited traveling-wave patterns that were remarkably stable and could persist for hours, provided that the flow and thermal boundary conditions remained unchanged. At intermediate flow rates, the flames transitioned to a steady ring-shaped flame anchored near the leading edge of the cavity. At elevated flow rates, local extinction/blow-off occurred as the flame front was displaced toward the rear side of the cavity. A series of parametric measurements over a wide range of equivalence ratios and flow rates were conducted, yielding a regime diagram of flame modes and their transition boundaries, as displayed in Fig. \ref{fig_04}.

A closer examination of the rotating flames revealed that the radial flow velocities at the cavity entrance were lower than the adiabatic flame speeds; however, flashback was suppressed by strong thermal quenching in the narrow channel upstream of the cavity, resulting in a dynamic balance between the local flame speed and the flow velocity. In the absence of the cavity, such a balance would be unstable inside the Hele-Shaw burner, since a negative perturbation of the local flame speed would push the flame front outward toward colder regions where the flame speed would be further reduced, while the velocity gradient in the channel would be insufficient to compensate for the change in flame speed. The leading edge of the cavity created a robust radial stabilization mechanism by introducing a region of strong negative velocity gradient via rapid flow expansion, and the expanded or retracted flame front induced by local perturbation could be trapped in this region. Meanwhile, there is no circumferential constraint on the flame, and the flame front can rotate in the angular direction of the cavity. A representative sequence of time-resolved OH* chemiluminescence images illustrating the transition from a steady ring-shaped flame to a rotating flame is provided in the Supplementary Material. \color{black}

\begin{figure}[h!]
\centering
\includegraphics[width=0.9\linewidth]{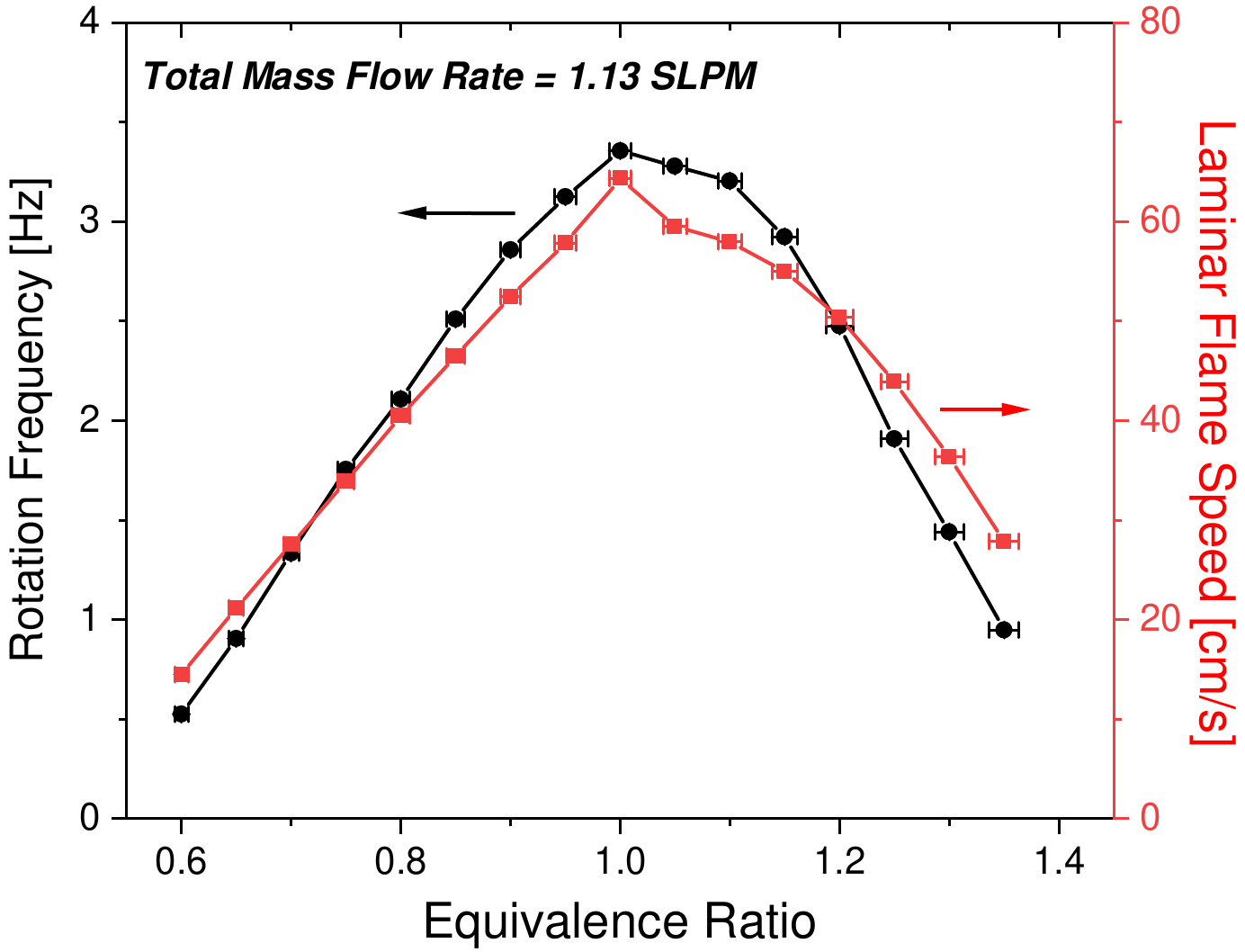}
\caption{\footnotesize The measured rotation frequency (black) and the calculated laminar flame speed (red) for single-headed rotating flames as functions of the equivalence ratio.}
\label{fig_05}
\end{figure}

At sufficiently low flow rates, the rotating flames contains a single front/wave. In this scenario, the projected area of the rotating flame along the cavity cross-section, $A_p$, can be estimated from the supply-consumption balance of the fresh mixture, $\dot{V} = 2\pi r_c f A_p$, where $r_c$ is the cavity leading-edge radius, and $f$ is the rotation frequency. In the present study, it was found that $f$ was roughly proportional to the reference adiabatic laminar flame speeds ($S_L$) evaluated at the measured surface temperatures (see Fig.\ref{fig_05}, calculated with Cantera \cite{cantera} using the FFCM-2 mechanism \cite{ZDV2023}), suggesting that for a fixed flow rate the flame fronts traveled in a nearly constant-shape fashion and the azimuthal propagation speed of the rotating head was controlled primarily by laminar flame propagation. For single-wave flames at different equivalence ratios shown in Fig. \ref{fig_05}, $A_p$ varies between 19 and 56 $mm^2$.

\begin{figure}[h!]
\centering
\includegraphics[width=\linewidth]{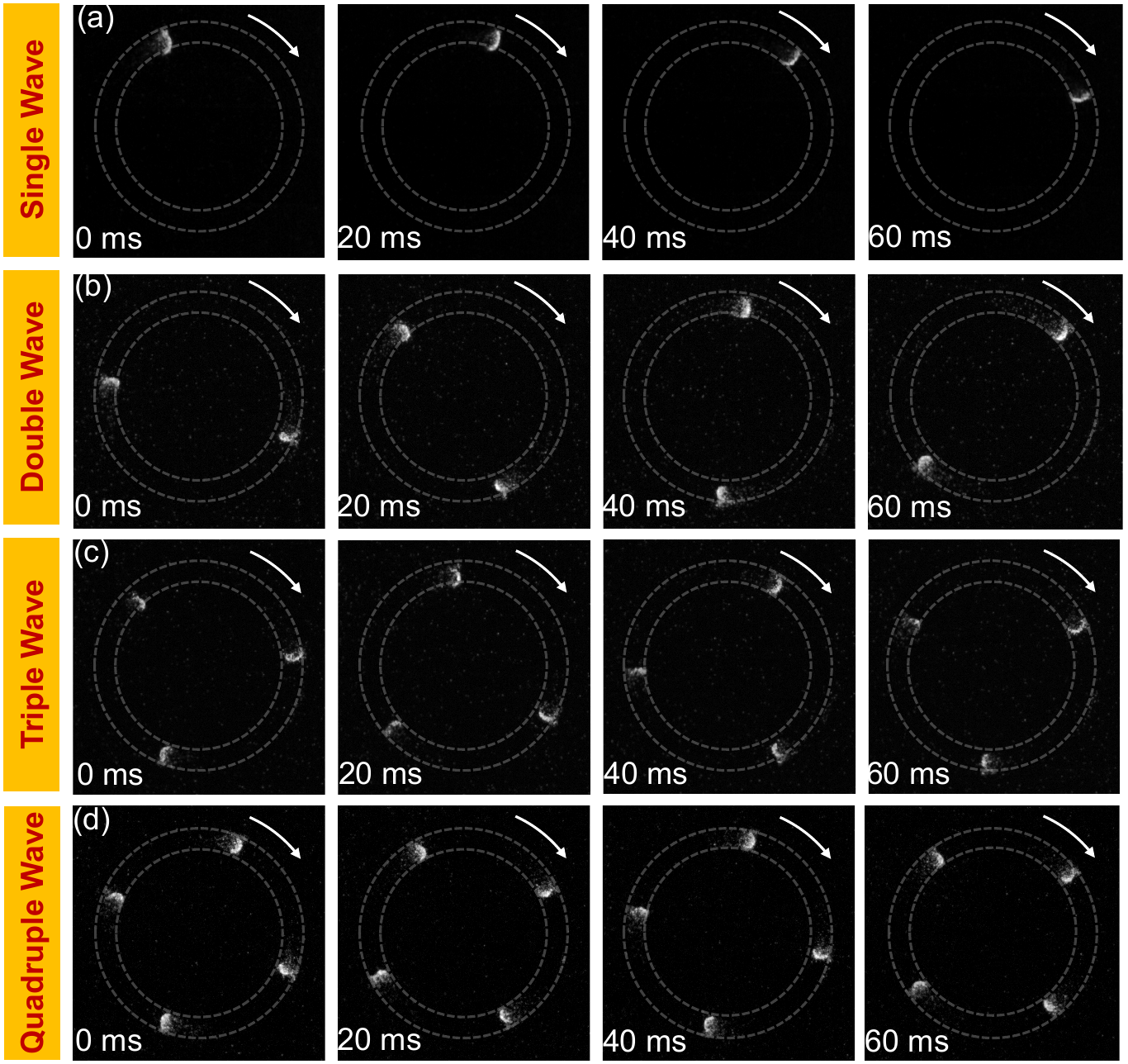}
\caption{\footnotesize Representative top-view OH* chemiluminescence images of single- and multi-wave rotating flames at $\phi$ = 1. The corresponding mass flow rates are: (a) 1.13 SLPM; (b) 4.52 SLPM; (c) 5.66 SLPM; (d) 6.80 SLPM. Dashed lines represent the locations of the cavity edges.}
\label{fig_06}
\end{figure}

For a single-wave rotating flame, the rotation frequency increases with the total flow rate when the flame front fully occupies the cavity cross-section, and the azimuthal propagation speed of the flame should match the laminar flame speed defined at the local thermal boundary conditions and equivalence ratio. At unit equivalence ratio and a representative surface temperature of 600 K on the high end, the laminar flame speed of CH$_4$-air mixture is approximately 1.4 m/s, which corresponds to a rotation frequency of 4.5 Hz. Based on this value, the maximum mass consumption rate of a single-wave rotating flame is estimated to be 3 SLPM. Therefore, at sufficiently high mass flow rates, the reactive mixture cannot be fully consumed by a single flame front. When the unburned mixture accumulates to a certain level, the rotating flames can split and evolve into multiple isolated fronts/waves of approximately equal spacing, as shown in Fig. \ref{fig_06}. A representative sequence of OH* chemiluminescence images illustrating the flame-splitting process can be found in the Supplementary Material.

For multi-wave rotating flames, the mixture supply-consumption balance is $\dot{V} = 2\pi N r_c f A_p$, with $N$ being the number of waves. The transitions between different $N$ exhibit a certain level of hysteresis and fuzziness (depend on the history of condition variations and are not exactly repeatable from run to run), but in general, $N$ is seen to increase with $\dot{V}$ in a superlinear fashion, suggesting that $A_p$ generally decreases with $N$. 

\begin{figure}[h!]
\centering
\includegraphics[width=0.9\linewidth]{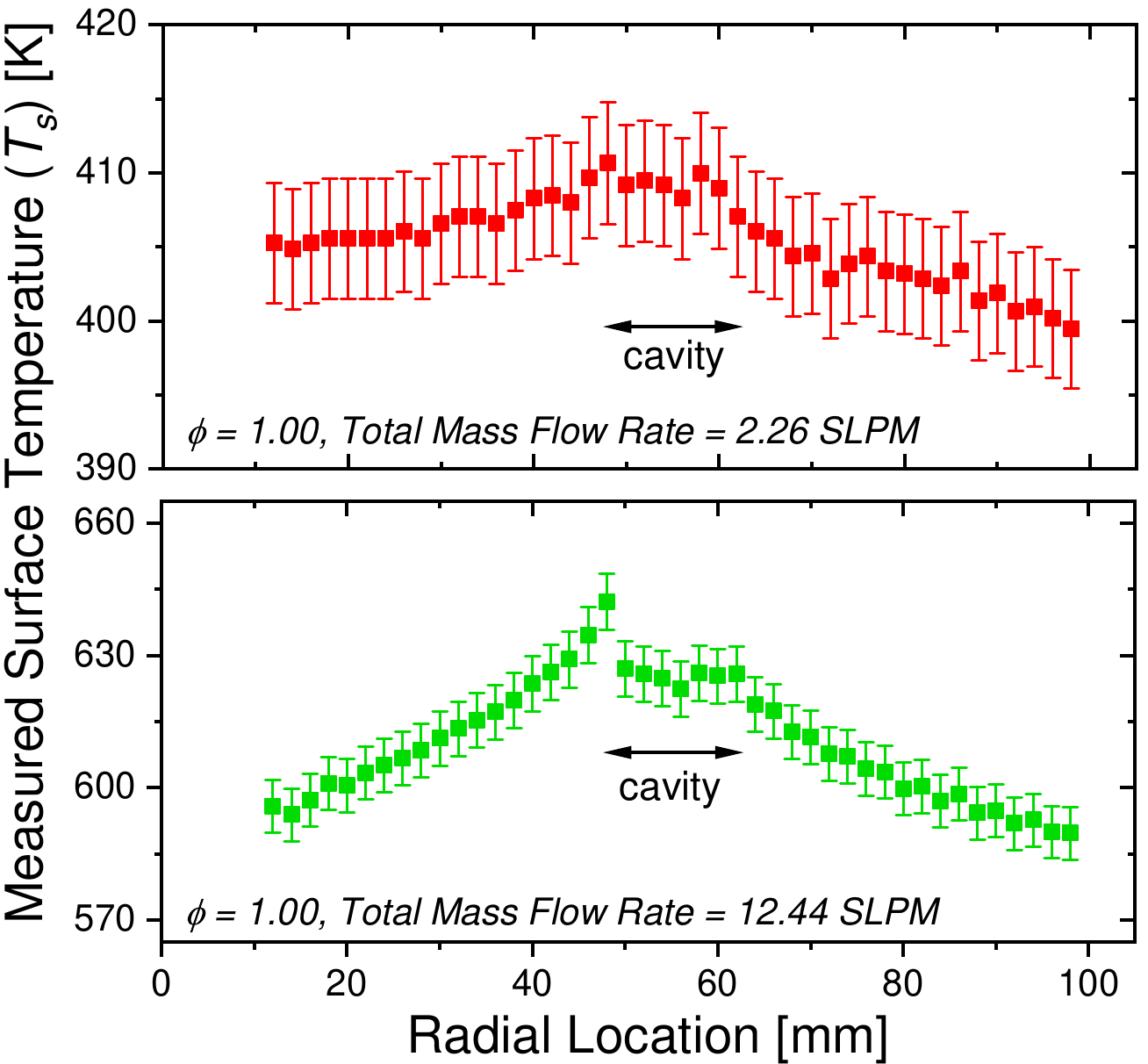}
\caption{\footnotesize Radial distributions of the steady-state surface temperature for representative cases of a single-headed rotating flame (top) and a ring-shaped steady flame (bottom).}
\label{fig_07}
\end{figure}

Rotating flames were routinely observed at total mass flow rates below 10 SLPM. At higher flow rates and equivalence ratios between 0.65 and 1.35, the flames transitioned to steady ring-shaped structures anchored near the leading edge of the cavity. The transition between rotating and steady flames can also be observed through the radial distribution of surface temperature, as shown in Fig. \ref{fig_07}. For a single-headed rotating flame, the surface temperature distribution was relatively smooth, indicating that the flame front spanned the entire width of the cavity. By contrast, for a typical steady ring-shaped flame, the surface temperature ($T_s$) reached its maximum value near the leading edge of the cavity and dropped abruptly inside the cavity, suggesting that the flame was highly concentrated near the cavity edge.

At elevated flow rates, local extinction/blow-off occurred as the flame front was pushed out of the rear side of the cavity, and total flame extinction was expected at sufficiently high flow rates. The transition flow rate varied significantly as a function of the equivalence ratio. Note that at equivalence ratios of $0.55\leq \phi \leq 0.60$ and $1.40\leq \phi \leq 1.45$, no steady ring-shaped flame was observed at any flow rate, likely due to the very low laminar flame speed and the pronounced influence of thermal quenching under these near-limit conditions. 

At very high flow rates, combustion in the cavity became turbulent, and flame front wrinkling, hot gas recirculation, and turbulent mixing with the unburnt gas further delayed flame blow-off. This was confirmed by a stoichiometric flame experiment conducted at the limit of the current mass flow controllers ($\dot{V} \approx$ 100 SLPM), where the flame was still sustained inside the burner. Further investigation of the blow-off limit is reserved for future studies.

\subsection{Complementary Numerical Results}
\addvspace{10pt}

\begin{figure}[h!]
\centering
\includegraphics[width=0.85\linewidth]{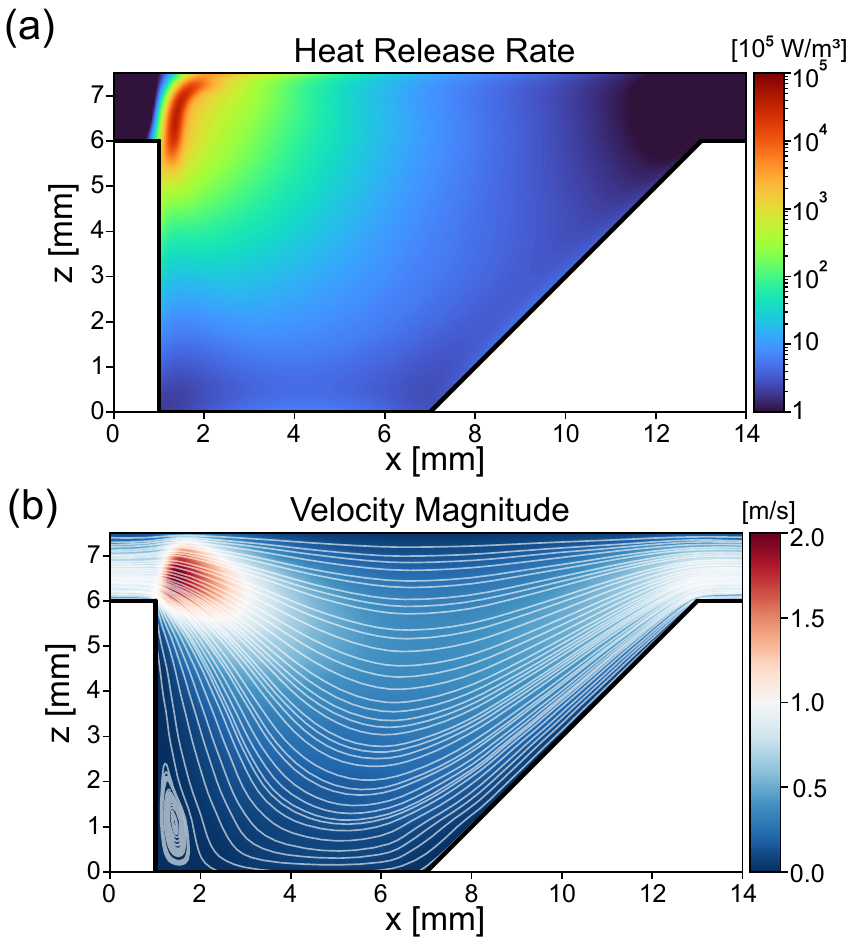}
\caption{\footnotesize 2-D distributions of (a) heat release rate $\dot{q}$ and (b) velocity magnitude and streamlines for a representative case near the transition boundary between rotating and steady flames. Conditions: $\phi$ = 1, $\dot{V}$ = 9.55 SLPM.}
\label{fig_08}
\end{figure}

Complementary numerical simulations were conducted to further examine the flame structure and flow field in the cavity under steady flame conditions near the mode transition boundaries. The results for a representative flame at conditions of $\phi$ = 1 and $\dot{V}$ = 9.55 SLPM is presented in Fig. \ref{fig_08}. The flame remains attached to the leading edge of the cavity and is stabilized in a low-speed region induced by sudden expansion of the flow. The flame front location appears relatively stable against small perturbations in the mass flow rate or laminar flame speed, due to a negative local velocity gradient manifested by the diverging streamlines ahead of the flame front. Additionally, flow recirculation near the bottom corners of the cavity contributes to flame stabilization by convecting hot products and reactive intermediates upstream.

\begin{figure}[h!]
\centering
\includegraphics[width=0.8\linewidth]{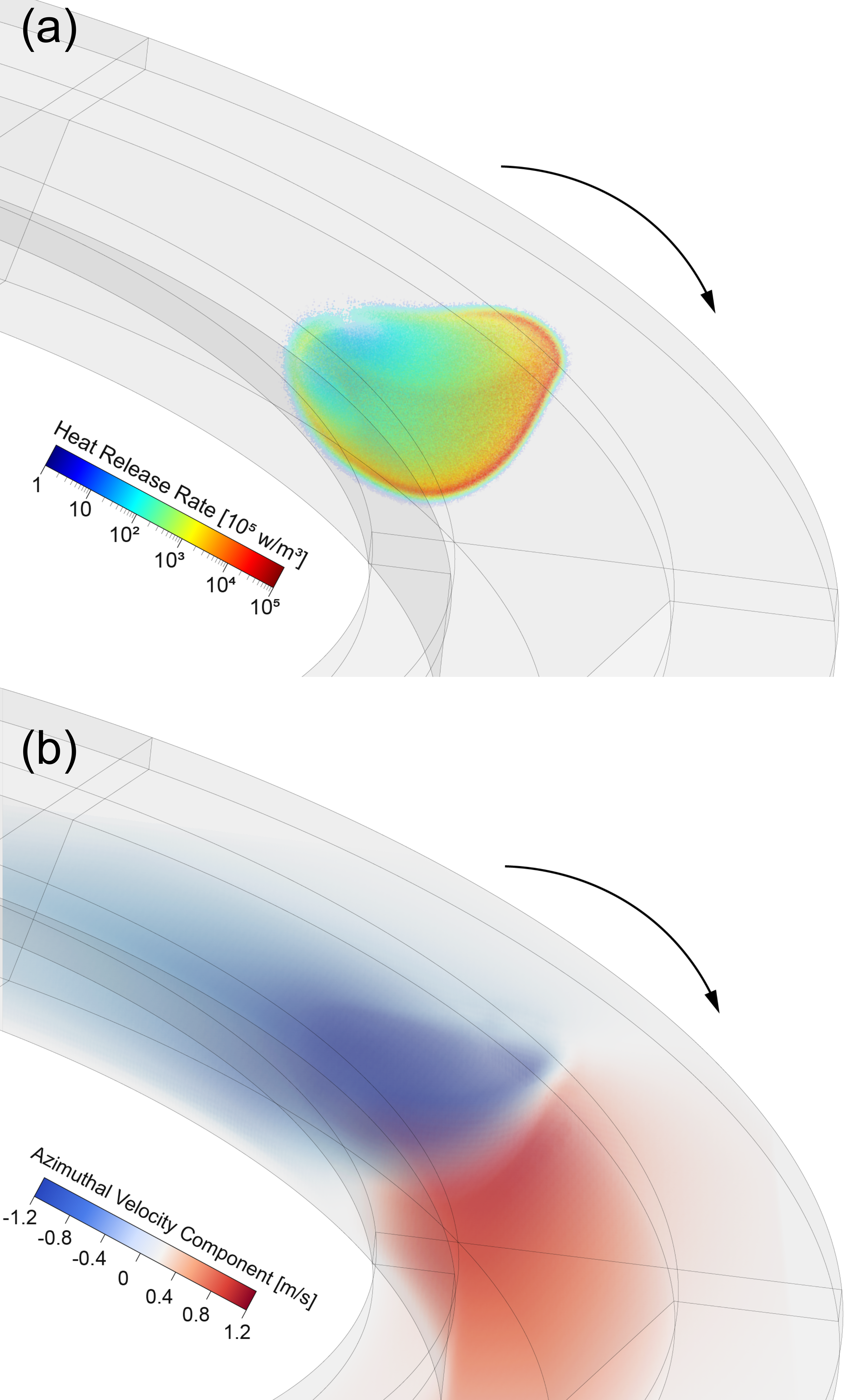}
\caption{\footnotesize 3-D structure of a rotating flame front at $\phi$ = 1 and $\dot{V}$ = 1.13 SLPM. (a) Heat release rate. (b) Azimuthal velocity component. Black arrows indicate the rotation direction.}
\label{fig_09}
\end{figure}

Fig. \ref{fig_09} shows the 3-D structure of a representative single-wave rotating flame, simulated under the same conditions as those in Fig. 6(a). The main purpose of this simulation is to illustrate the spatial distributions of heat-release rate and azimuthal velocity across the rotating flame front. From the spatial distribution of the heat release rate, it is seen that the rotating flame front is mostly constrained within the cavity. Across the flame front, the azimuthal velocity (evaluated in a fixed reference frame) changes sign due to thermal expansion. Further analysis shows that the projected propagation speed of the flame front along its normal direction is of the same order as the laminar flame speed calculated with Cantera, and the extracted rotation frequency is 3.12 Hz, which agrees reasonably well with the measured value (3.40 Hz).

\subsection{Mode Transition Boundaries}
\addvspace{10pt}

Two distinct types of mode transitions boundaries -- one between the rotating flames and steady ring-shaped flames, and the other between steady flames and local extinction -- are analyzed.

A key observation from Fig. \ref{fig_04} is that the rotating-to-steady transition boundary exhibits little dependence on the equivalence ratio. For methane-air mixtures at a gap distance of 1.5 mm, the transition flow rate is nearly constant ($\sim$ 10 SLPM), while the reference laminar flame speed (evaluated at the measured surface temperature) varies between 0.72 and 1.35 m/s. 

A plausible explanation for the underlying mechanism of the rotating-to-steady transition is postulated, based on a physical analogy to previous studies of flame-wall interaction in confined premixed flames \cite{dreizler2015advanced, sereshchenko2011formation, minaev2009splitting, fan2009experimental}. Near the leading edge of the annular cavity, a steady ring flame propagates approximately parallel to the neighboring wall. The flame tends to flashback as the low flow rate decreases, but this flashback tendency is counteracted by wall quenching (which reduces the flame speed) until a point where the flame speed is too low to be stable. At this point, upon small perturbations the ring-shaped flame can switch to a mode of lower loss and higher flame speed, i.e., the rotating flame. The critical flow rate of mode transition can be estimated based on a non-dimensional analysis similar to previous studies of sidewall quenching \cite{von1953thermal, lu1991unsteady}, as follows.

For a steady ring-shaped flame, the mass flow rate of fresh reactant per unit length along the circumference of the cavity is $\dot{V} /2\pi r_c $, from which an effective flame height can be calculated by dividing this quantity by the laminar flame speed, i.e., $h = {\dot{V}}/{2\pi r_c S_L}$. In the absence of external heating, $h$ must be significantly larger than the diffusion thickness of the flame, $\delta = {\alpha_u}/{S_L}$, and their ratio defines a Peclet number representing the relative strength of advective transport to thermal diffusion: $Pe = h/\delta = \dot{V} /2\pi r_c \alpha_u$. This yields a dimensionless criterion for the rotating-to-steady flame transition. For the transition boundary shown in Fig. \ref{fig_04}, the corresponding Peclet number is $Pe =7.2~\pm~1.3$, which in close agreement with the critical Peclet number for sidewall quenching ($Pe \sim$ 7 as reported in \cite{poinsot2005theoretical, dreizler2015advanced}). 

\begin{figure}[h]
\centering
\includegraphics[width=0.9\linewidth]{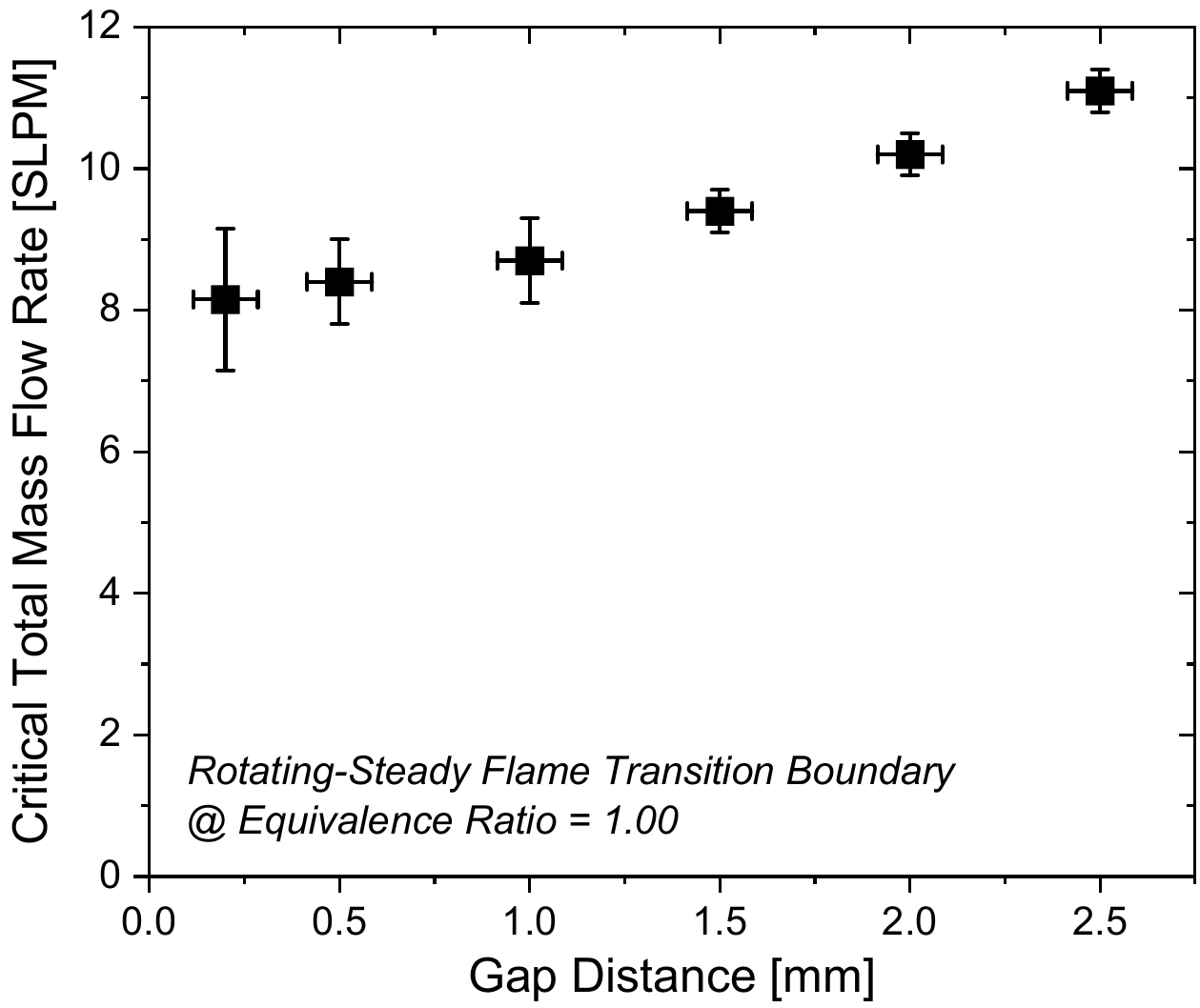}
\caption{\footnotesize Critical total mass flow rate at the mode transition boundary between rotating and steady flames. Results are shown for stoichiometric mixtures at various gap distances.}
\label{fig_10}
\end{figure}

The critical total mass flow rate at the mode transition boundary is shown as a function of the gap distance in Fig. \ref{fig_10}. Although the average velocity at the cavity entrance varies by nearly an order of magnitude, the critical total mass flow rate increases by no more than 50\% as the gap distance changes from 0.2 mm to 2.5 mm. As the gap distance approaches zero, the critical total mass flow rate tends toward an asymptotic value of approximately 8 SLPM. Data were not collected at gap distances greater than 2.5 mm due to local flashback upstream of the cavity entrance. The insensitivity of the transition flow rate to the gap distance is consistent with the previous analysis based on the critical Peclet number, which does not depend on the gap distance.

\begin{figure}[h]
\centering
\includegraphics[width=0.9\linewidth]{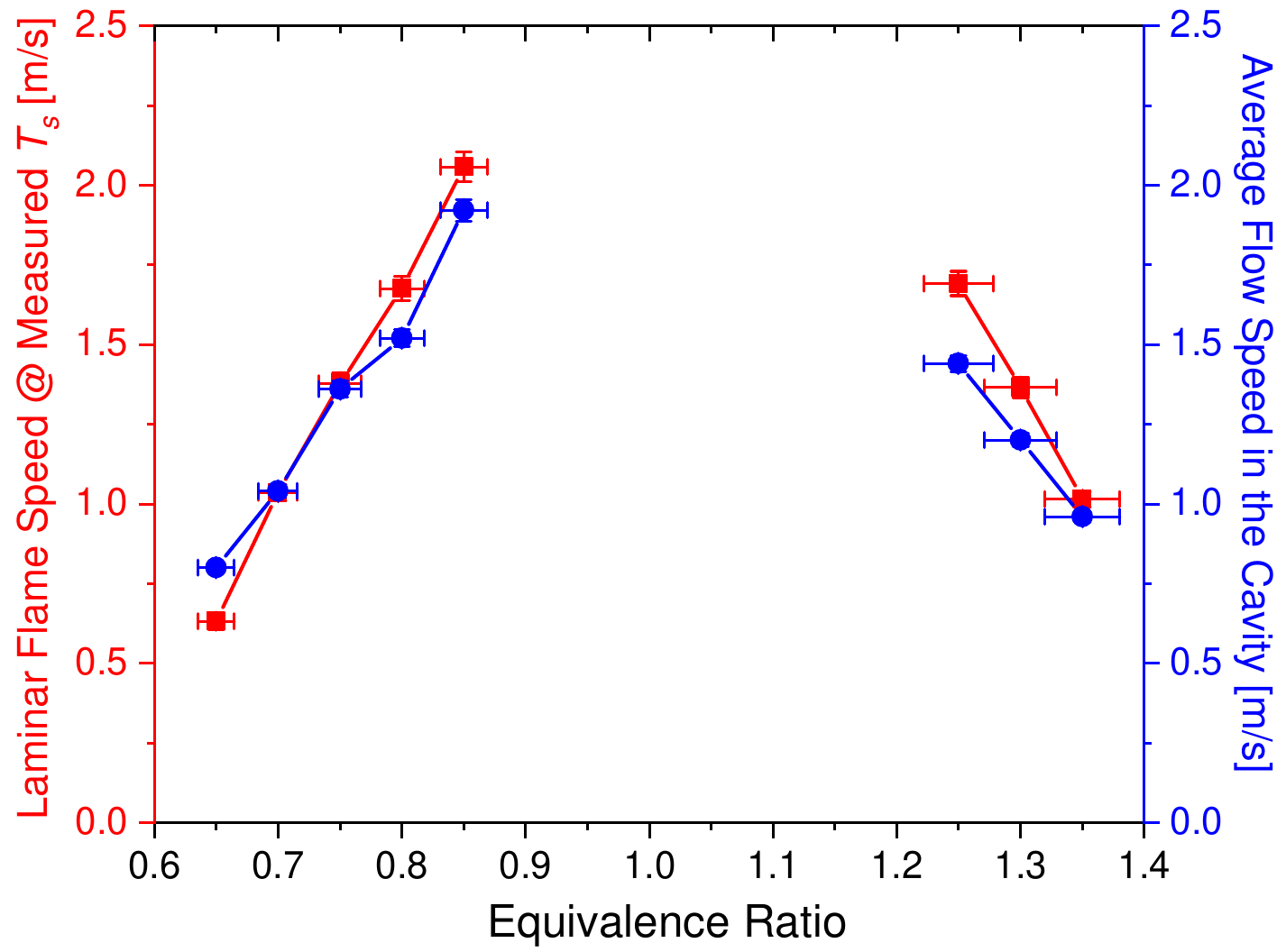}
\caption{\footnotesize Characteristic speeds at the transition boundary between steady flames and local extinction.}
\label{fig_11}
\end{figure}

Unlike the rotating-steady flame transition boundary, the transition boundary between steady flames and local extinction is strongly dependent on the equivalence ratio (see Fig. \ref{fig_04}). The characteristic flow velocity (defined as the radial flow velocity at the cavity leading edge), as shown in Fig. \ref{fig_11}, is of the same order as, and exhibits a similar equivalence-ratio dependence to, the reference laminar flame speed evaluated at the measured surface temperature. Local blow-off occurs when the average flow velocity exceeds the laminar flame propagation speed. The corresponding critical mass flow rate is insensitive to the gap distance, as the cavity depth is significantly larger.

\subsection{Experiments with Other Fuels}
\addvspace{10pt}

\begin{figure}[h!]
\centering
\includegraphics[width=0.9\linewidth]{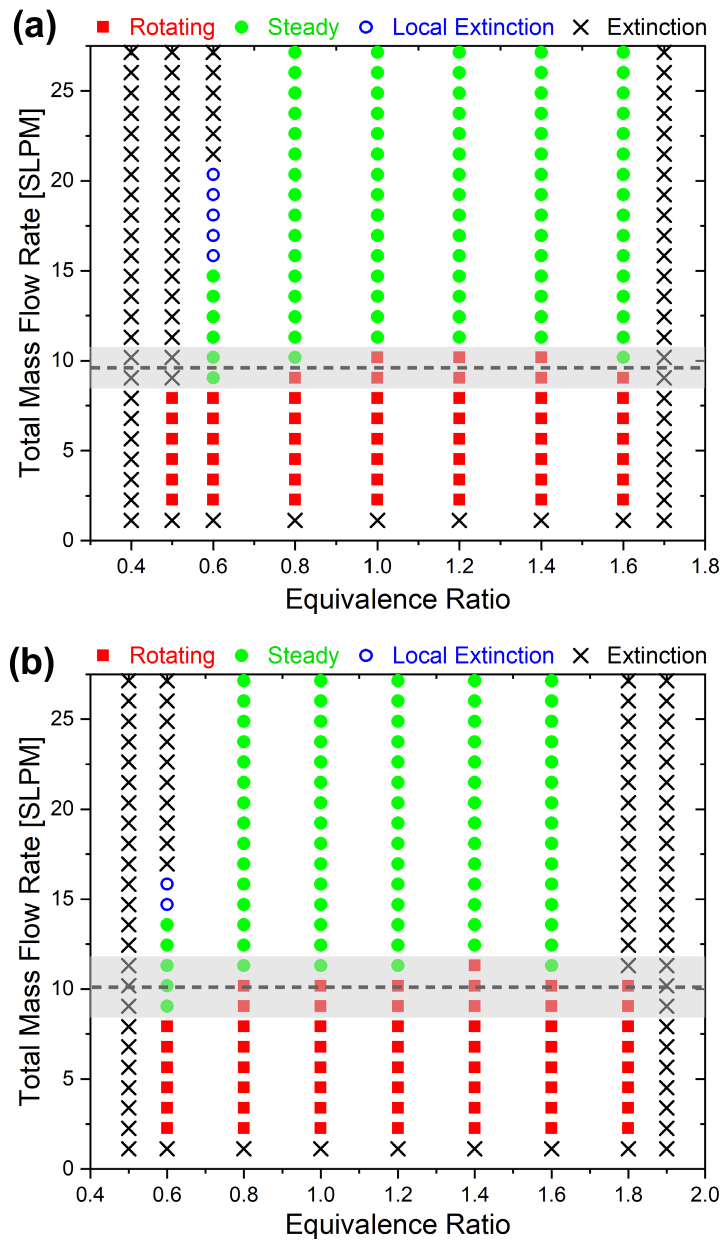}
\caption{\footnotesize Regime diagrams of the experimentally observed flame modes for (a) C$_3$H$_8$-air and (b) DME-air mixtures in the cavity-stabilized Hele-Shaw burner. Measurements were conducted at a gap distance of 1.5 mm.}
\label{fig_12}
\end{figure}

Additional experiments were conducted with other fuels, including propane (C$_3$H$_8$) and dimethyl ether (DME). Similar patterns of rotating flames were observed at total mass flow rates below 10 SLPM. The regime diagrams of flame modes for these fuels are presented in Fig. \ref{fig_12}. Compared with CH$_4$-air rotating flames, the equivalence-ratio ranges for rotating flames are wider for C$_3$H$_8$-air (0.50 - 1.60) and DME-air (0.60 - 1.80), especially on the rich side. This behavior is consistent with the general trend of dependence on the equivalence-ratio of the laminar flame speed; under rich conditions, C$_3$H$_8$ and DME exhibit higher laminar flame speeds than CH$_4$. In addition, compared to CH$_4$ flames, the local-extinction regime is much narrower for C$_3$H$_8$ and DME flames, likely due to the higher surface temperatures associated with fuels of greater heating value.

Despite the differences in equivalence ratio ranges, the critical total mass flow rates at the rotating-steady transition boundary for other hydrocarbon flames, e.g. C$_3$H$_8$-air flames (9.6 $\pm$ 1.1 SLPM,  $Pe=7.2~\pm~1.3$) and DME-air flames (10.2 $\pm$ 1.7 SLPM,  $Pe=7.9~\pm~1.0$)  are very close to that for CH$_4$-air flames (9.6 $\pm$ 1.1 SLPM, $Pe=7.8~\pm~1.3$). The original $Pe$ number data for each fuel, along with the associated mixture temperatures and thermal diffusivities, are documented in the Supplementary Materials. Further investigation of the transition boundary for fuels with markedly different transport and flame properties, such as hydrogen (H$_2$) and ammonia (NH$_3$), is warranted in future studies. 

\section{Conclusions}
\addvspace{10pt}

Self-organized rotating flames of CH$_4$–air mixtures were routinely observed at equivalence ratios between 0.55 and 1.45 in an open circular Hele-Shaw burner with an annulus cavity flame holder. These rotating flame patterns demonstrated notable robustness against perturbations in flow conditions, adapting over a wide range of laminar flame speeds and flow velocities by self-adjusting their rotation frequency and the number of waves. The transitions of these rotating flames to steady ring-shaped flames and to flames of local extinction were studied at elevated flow rates, yielding a regime diagram that details the various flame modes and their transition boundaries. Additional experiments with propane (C$_3$H$_8$) and dimethyl ether (DME) fuels showed similar rotating-flame patterns at total mass flow rates below 10 SLPM. The critical total mass flow rate at the rotating-steady flame mode transition boundary (approximately 10 SLPM) was found to be relatively insensitive to equivalence ratio, gap distance, and fuel type. A nondimensional analysis indicates that the rotating-to-steady transition is governed by Peclet numbers for the present geometry and the tested fuels. A physical explanation for the mode transition is also postulated. These findings are useful to both the fundamental understanding of flame dynamics in micro-channels and the practical applications of micro‑combustors and micro‑combustion technologies.

\acknowledgement{CRediT authorship contribution statement} 
\addvspace{10pt}

\textbf{Xiangyu Nie}: Investigation; Formal analysis; Writing -- original draft. \textbf{Shengkai Wang}: Conceptualization; Methodology; Resource; Supervision; Writing -- original draft, review \& editing.

\acknowledgement{Declaration of competing interest} 
\addvspace{10pt}

The authors declare that they have no known competing financial interests or personal relationships that could have appeared to influence the work reported in this paper.

\acknowledgement{Acknowledgments}
\addvspace{10pt}

This work was supported by the National Key Research and Development Program of China under Grant No. 2025YFF0511801 and by the National Natural Science Foundation of China under Grants No. 12472278 and No. 92152108. Numerical simulations were supported by the High-Performance Computing Platform of Peking University.

% -------------------------------------------------------------------- %
% -------------------------------------------------------------------- %
% -------------------------------------------------------------------- %
\footnotesize
\baselineskip 9pt

% -------------------------------------------------------------------- %
% -------------------------------------------------------------------- %
% -------------------------------------------------------------------- %
\clearpage
\thispagestyle{empty}
\bibliographystyle{proci}
\bibliography{References}

% -------------------------------------------------------------------- %
% -------------------------------------------------------------------- %
% -------------------------------------------------------------------- %

\newpage

\small
\baselineskip 10pt

% -------------------------------------------------------------------- %
% -------------------------------------------------------------------- %
% -------------------------------------------------------------------- %

\end{document}